\documentclass[pra,twocolumn,superscriptaddress]{revtex4}
\usepackage{epsfig,amsfonts}
\begin{document}
\title{Collisions of Slow Highly Charged Ions with Surfaces}
\author{J. Burgd\"orfer}
\affiliation{Institute for Theoretical Physics, Vienna University of Technology, A-1040 Vienna, Austria, EU}
\author{C. Lemell}
\affiliation{Institute for Theoretical Physics, Vienna University of Technology, A-1040 Vienna, Austria, EU}
\author{K. Schiessl}
\affiliation{Institute for Theoretical Physics, Vienna University of Technology, A-1040 Vienna, Austria, EU}
\author{B. Solleder}
\affiliation{Institute for Theoretical Physics, Vienna University of Technology, A-1040 Vienna, Austria, EU}
\author{C. Reinhold}
\affiliation{Oak Ridge National Laboratory, Oak Ridge, TN 37831, USA}
\author{K. T\H OK\'ESI}
\affiliation{Institute for Theoretical Physics, Vienna University of Technology, A-1040 Vienna, Austria, EU\\
Institute of Nuclear Research of the Hungarian Academy of Sciences (ATOMKI),
H-4001 Debrecen, P.O.Box 51, Hungary, EU}
\author{L. Wirtz}
\affiliation{Institute for Electronics, Microelectronics and Nanotechnology, B.P. 60069, 59692 Villeneuve d'Ascq Cedex, France, EU}
\begin{abstract}
Progress in the study of collisions of multiply charged ions with surfaces is reviewed with the help of a few recent examples. They range from fundamental quasi-one electron processes to highly complex ablation and material modification processes. Open questions and possible future directions will be discussed.
\end{abstract}
\maketitle
\section{Introduction}
With the advent of highly charged $(Q \gg 1)$, low energy ion sources, the interaction of slow multiply charged ions with surfaces has developed into one of the most active areas in the field of particle-solid interactions. On the most fundamental level, its interest is derived from the complex many-body response of metal electrons to the strong Coulomb perturbation characterized by a large Sommerfeld parameter $\eta = Q/\upsilon_p \gg 1$. The neutralization is a true multi-electron capture (and loss) process involving up to the order of $\approx 100$ electrons and posing a considerable challenge to theory. Furthermore, resonant transfer processes involve highly excited levels in the ions and are expected to set in at large distances from the surfaces when the Rydberg wavefunction begins to ``touch'' the surface. The study of multiply and highly charged ion-surface interaction is also of considerable importance for the understanding of surface damage and plasma-wall interactions. 

As experimental techniques have become more sophisticated, the motivation for these studies has broadened to elucidate the dynamical response of the many-body system of the surface to a very strong Coulomb perturbation on one side and to material science, surface diagnostics, characterization and modification on the other. Given the diversity of the subject and its rapid growth, no attempt of completeness of this overview will be made.

We will briefly review the neutralization scenario now widely accepted and discuss new experimental and theoretical results which have expanded our understanding of highly charged ion (HCI) surface interactions. The key element will be the ``classical over barrier'' (COB) model for electron transfer into Rydberg states \cite{burg91,burg93,hagg}. This model has been remarkably successful in a qualitative and, sometimes, quantitative description of charge transfer between surfaces and HCI. Originally developed for metals, it has been extended to insulators. Despite its success, its conceptual foundation and range of validity is far from being well established.

Only for simple, effective one-electron processes, fully quantum mechanical models for electronic processes have been developed, a few of which will be briefly discussed. First attempts to develop a many-body theory have been taken within the framework of time-dependent density functional theory (TDDFT). Both the potential as well as major limitations of such methods will be summarized.

Driven by both conceptual interest and by technical applications, nano-scale surfaces have taken center stage. Interactions with internal walls of both metallic and insulating nano-capillaries will be considered. We will focus on material science applications of collisions with surfaces. Here, the use of ion scattering as a tool to study surface magnetism and for active manipulation and modification of surfaces by charged particle impact are currently hot topics of investigation. The latter holds the promise of self-organized and, possibly functionally controlled, nanostructures. This paper is not meant as a comprehensive review but as an introduction into current developments and unsolved problems in the field of ion-surface scattering. More detailed informations can be found in the list of references and in recent reviews \cite{burg1993,wint,winter,arna}. 

\section{The Collision Scenario}
\begin{figure}[t]
\begin{center}
\epsfig{file=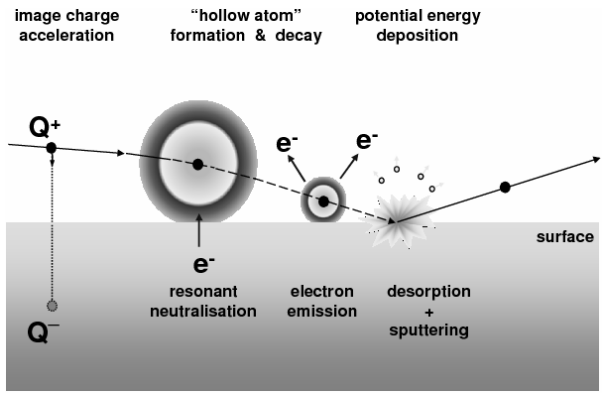,width=8cm,clip=}
%\epsfxsize=10cm   %width of figure - will enlarge/reduce the figures
%\epsfbox{fig1111.eps}
\end{center}
%\figurebox{3cm}{4cm}{}
\caption{
\label{fig:1} Scenario for the interaction of a slow HCI of charge $Q$ with surfaces (left to right): attraction by image force, multiple electron capture into high Rydberg states and formation of hollow atoms, dissipation of potential energy by Auger deexcitation, autoionization, and photoemission, relaxation near point of impact with ablation of surface particles, electrons, and heating, finally reflexion or penetration of projectile (Fig.\ from \protect\cite{fritznim}).}
\end{figure}

Interaction of slow highly charged ions (HCI) with surfaces is a true many-body problem involving a complex array of processes on different time and energy scales, schematically depicted in Fig.\ \ref{fig:1}.
They are driven by the large amount of potential energy (PE) given by the sum over all successive ionization potential $(I_p)_i$ of the HCI,
\begin{equation}
\label{eq:1}
PE = \sum_{i=1}^{Q} (I_p)_i \, ,
\end{equation}
brought into the collision. PE can easily reach tens $keV's$ and, for slow ions, considerably exceed the kinetic energy $\frac{1}{2} M v_p^2 $ of the projectile. For grazing incidence where the normal component $v_{\perp}$ is small compared to the parallel component, $v_{\perp} \ll v_{\parallel}$, we have 
\begin{equation}
\label{eq:2}
\frac{1}{2} M v_\perp \ll PE
\end{equation}
such that penetration of the projectile into deeper layers of the solid can be ruled out. 

The ensuing interaction scenario is dominated by competing mechanisms for dynamical screening, conversion and dissipation of the incident energy into electronic and nuclear degrees of freedom, particle emission and, ultimately structural deformations of the solid surface. In the first stage of this scenario, the ion polarizes the surface and creates a (self) image potential, $V^{SI}$ 
\begin{equation}
\label{eq:3}
V^{SI} (R)= -\frac{Q^2}{4R} \frac{\left(\varepsilon(\omega)-1\right)}{\left(\varepsilon(\omega)+1\right)} \, ,
\end{equation}
where $\varepsilon (\omega)$ is the bulk dielectric function of the target while $(\varepsilon -1)/(\varepsilon + 1)$
represent the optical limit $ Q \rightarrow 0$ of the surface dielectric function $\varepsilon (\omega, Q)$ \cite{burg1993}. Ideal conductors and thus metallic surfaces correspond to the limit $\varepsilon (\omega) \gg 1$ for $\omega \ll 1$. The attractive image force leads to an acceleration, and thus to a bending of the trajectory towards the surface. At a critical distance $R = R_c$ the electronic response is no longer confined to a polarization within the surface but results in transfer of electrons into high lying states of the projectile. The formation of ``hollow atoms'' \cite{briand} (more precisely, hollow ions) marks this second stage. As the projectile approaches the surface, part of the potential energy is released by electron emission in Auger processes involving both electrons localized around the ion as well as near the surface. The relaxation during this stage is, however, too slow to render the projectile in its ground state. In vicinity to the surface, quasi-resonant capture from inner shells (``side feeding'' \cite{folk}) as well as two-center Auger capture \cite{hag} begin to fill inner shells marking the fourth stage of this process. The key feature of the slow deexcitation is that a significant fraction of the incident potential energy remains to be relaxed as the ion hits the surface. Consequently, dissipation will involve the electronic degrees of freedom of the topmost layers and, by way of electron-phonon coupling, of atoms and ions near the surface. The final stage is then the creation of defects, heating of the lattice, and ablation of surface material (``potential sputtering'' \cite{par,neid}).\\
Developing a quantitative description with predictive power for such a complex array of processes has remained a challenge. Nevertheless, some progress made to date will be highlighted in the following.

\section{Theoretical Methods}
\subsection{Classical Over Barrier Model}
Classical models for the electronic dynamics have a long-standing tradition in atomic collision physics going back to Bohr's account of energy loss of charged particles \cite{bohr48}. This tradition was continued with Thomas' genius analysis of electron capture in 1927 \cite{thom} and later with the binary encounter model for ionization \cite{bonsen}. For problems with one active electron, the ``classical trajectory Monte Carlo (CTMC) Method'' \cite{abrines} was developed as an efficient algorithm with, quite often, quantitative predictive power. Classical methods are appealing in view of their simplicity compared to quantum calculations, even for simple systems. While for fundamental atomic collision process, such as one-electron ionization and capture, quantum ab initio calculations are possible \cite{wint94,min}, resorting to a classical descriptions is almost in inevitable for complex many-body systems such as collisions with surfaces. Among classical models, the classical over-barrier (COB) model has proven to be versatile and remarkably successful. 

The COB model was originally developed for one-electron capture into highly charged ions in ion-atom collisions by Ryufuku et al. \cite{ryu} based on earlier work by Bohr and Lindhard \cite{bohr} and later extended by Barany et al. \cite{bar} and Niehaus \cite{niehaus} to incorporate multi-electron transfer. Its extension to ion-surface collisions \cite{burg91,burg93,hagg} provides a simple framework for the description of ion-surface interactions. The physical significance of the COB model is derived from the fact that only classically allowed over-the-barrier processes as opposed to tunneling are sufficiently fast to be effective within the characteristic interaction time of the ion with the surface.\\
An 'active' electron crossing the barrier is subject to the potential $V(\vec{r},\vec{R})$,
\begin{equation}
\label{eq:4}
V(\vec{r},\vec{R})=V_e(\vec{r})+V^I_{pe}(\vec{r},\vec{R})+V_{pe}(\vec{r},\vec{R})
\end{equation}
where $\vec{r}=(x,y,z)$ is the position of the electron, with $x$ and $y$ parallel to the surface and $z$ perpendicular to the surface, and $\vec{R}=(R_x,R_y,R_z)$ is the position of the projectile (an HCI). The direct interaction between the electron and the HCI, $V_{pe}$, is given for hydrogenic projectiles by
\begin{equation}
\label{eq:5}
V_{pe}(\vec{r},\vec{R})=-\frac{Q}{|\vec{r}-\vec{R}|}\, .
\end{equation}
Corresponding single-particle core potentials for non-hydrogenic ions are available in the literature \cite{gar}. The interaction potential between the electron and the surface in absence of the HCI is denoted by $V_e$. The ``image potential'' due to the presence of the HCI is denoted by $V_{pe}^{I}$. The latter two potentials are markedly different for metals and for insulators \cite{hagg,jenn,deutscher97}. Choosing for simplicity pure Coulomb potentials the electronic self-image potential (Eq. (\ref{eq:3} )) as the large-$z$ limit of $V_e$ for metals, the position of the saddle point $z_s, dV(z_s)d_z=0$ and the barrier height $V(z_s, R)$ can be determined analytically. The critical distance follows from \cite{burg91,burg93,hagg,burg1993}, 
\begin{equation}
\label{eq:6}
V(z_s, R_c)=E_T=-W \, ,
\end{equation}
where $W$ is the work function of the surface, as 
\begin{equation}
\label{eq:7}
R_c(Q)\simeq \frac{\sqrt{2Q}}{W} \, .
\end{equation}
The COB model predicts, furthermore, the critical quantum number $n_c$ into which the first resonant capture event takes place,
\begin{equation}
\label{eq:8}
n_c=\frac{Q}{\sqrt{2W}} \left( \frac{1}{1+{\frac{Q-1/2}{\sqrt{8Q}}}} \right)^{1/2} \, .
\end{equation}
\begin{figure}[t]
\begin{center}
\epsfxsize=8cm   %width of figure - will enlarge/reduce the figures
\epsfbox{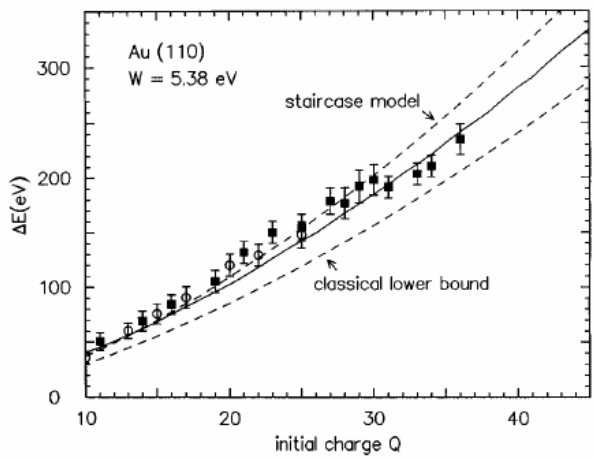}
\end{center}
\caption{
\label{fig:2}
Scaled energy gain $\Delta E/W$ due to image acceleration for different targets. Experimental data: $O: A1; \Delta, \square, $Au(110),$ \bullet,$ Au. Also shown are the COB model (see Eq. \ref{eq:9}) and the classical lower bound, from Ref. 28 and refs. therein.}
\end{figure}
These predictions pertain to the first stages of the neutralization scenario and were initially difficult to verify since the subsequent violent relaxation processes tend to erase the memory on earlier stages. Eq. (\ref{eq:7}) was first indirectly tested (Fig.~\ref{fig:2}) by measurements \cite{wint93,aumayr93} of the energy gain due to the image acceleration (Eq. (\ref{eq:3})).
Since the charge of the projectiles is reduced by one unit $(Q'=Q,Q-1 ...)$ at each successive critical distance, $R_c(Q')$ (Eq. (\ref{eq:7})), the energy gain along the neutralization sequence (``staircase'') follows as [2]
\begin{equation}
\label{eq:9}
\Delta E=\frac{W}{3\sqrt{2}} Q^{3/2}
\end{equation}
in agreement with experimental data \cite{wint93,aumayr93,lem}. Similar estimates can be derived for insulators [3].\\
Direct test became available with the help of novel targets, metallic nanocapillaries, pioneered by the group of Yamazaki et al. \cite{nino}. Interaction with the internal walls of capillaries allows to select trajectories that avoid a close encounter with the surface (trajectories of type 3 (Fig.~\ref{fig:3}))
\begin{figure}[t]
\begin{center}
\epsfxsize=8cm   %width of figure - will enlarge/reduce the figures
\epsfbox{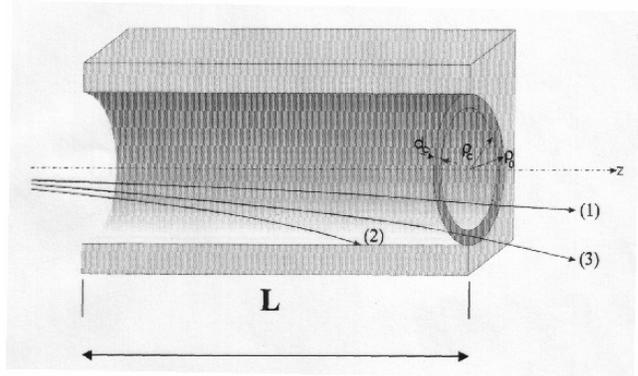}
\end{center}
\caption{
\label{fig:3}
Sketch of nanocapillary and typical ion trajectories (aspect ratio not to scale), trajectories leaving the capillary in its initial charge state (type 1); trajectories undergoing grazing incidence scattering (type 2), and trajectories leaving capillary before touching down probing early stages of hollow-atom formation (type 3).}
\end{figure}
thus allowing spectroscopic as well as charge-state analysis of the early stages of the neutralization scenario. Application of the COB allows the determination of the charge state fractions proportional to the areas of concentric rings whose thicknesses are determined by Eq. (\ref{eq:7}). Good agreement of the COB simulations \cite{tok}, even when relaxation processes are neglected clearly supports the validity of Eq. (\ref{eq:7}) (Fig.~\ref{fig:4}).
\begin{figure}[t]
\begin{center}
\epsfxsize=7cm   %width of figure - will enlarge/reduce the figures
\epsfbox{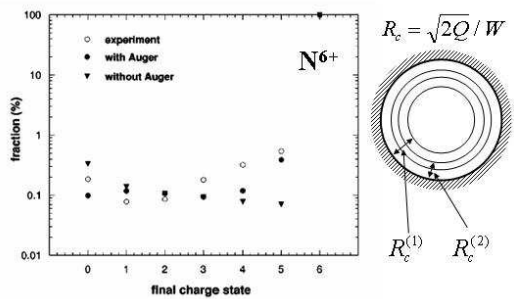}
\end{center}
\caption{
\label{fig:4}
left: Final state charge fraction of $N^{6+}$ penetrating $Ni$ nanocapillary.  Open circles: exp. (ref. 29), triangle: prediction by geometric cross sections (right), full circle: COB simulation (ref. 30), right: Decomposition of cross section into rings corresponding to sequential capture of electrons at the critical distance for each charge state.}
\end{figure}
Very recently, much more detailed evidence on the early stages of hollow atom formation became available through spectroscopy of optical emission from high Rydberg states in $Ar^{Q} (Q=8-12)$. From the line intensity along the transitions $n \rightarrow n -1$, the population of $n$ shells could be determined \cite{mor} (Fig.~\ref{fig:5}). The peak in the initial state occupation follows the prediction (Eq. \ref{eq:8}). 
\begin{figure}[t]
\begin{center}
\epsfxsize=8cm   %width of figure - will enlarge/reduce the figures
\epsfbox{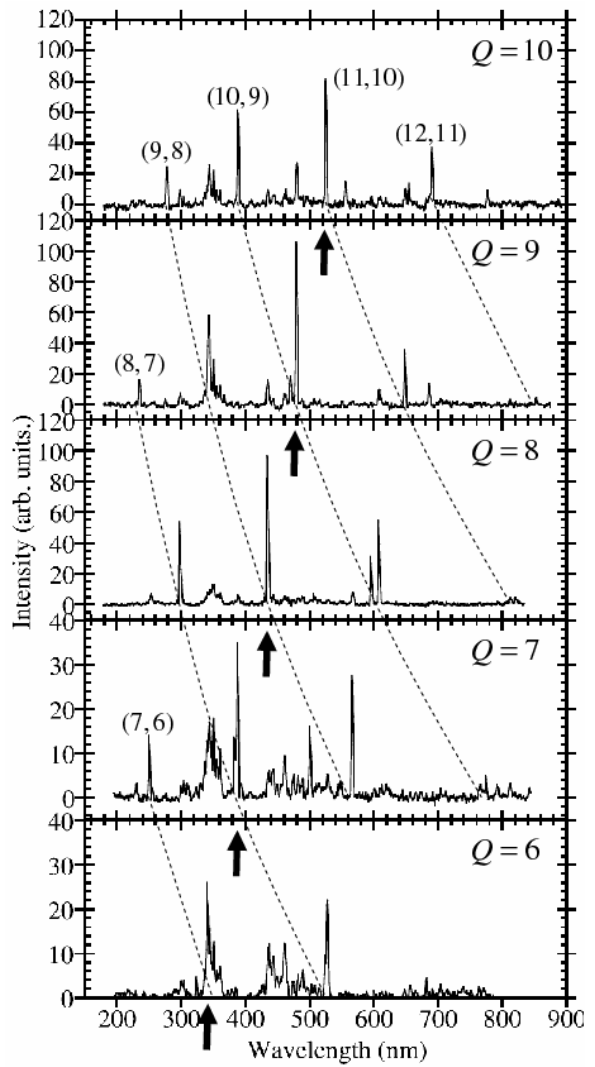}
\end{center}
\caption{
\label{fig:5}
Optical emission spectra for 2 keV/amu incident Ar$^{q+}$ ions. Lines connected by the solid lines are attributed to $\Delta n=1$ transitions of Ar$^{(q-1)+}$ (from ref. 31). Predictions for $n_c$ (Eq. (8)) marked by arrows.}
\end{figure}

\subsection{Quantum Mechanical Foundations}
\label{quantum}
The COB model hinges on several approximations. The most drastic ones include: tunneling (i.e. below barrier) transitions are neglected since they have small rates compared to over-barrier processes so that they give only a minor contribution to the reaction rate. Atomic states are assumed to exist as resonances near and above the barrier when the electron is classically allowed to escape. Optical dipole transitions within the projectile driven by dynamical polarization fields in the target (e.g. plasmon excitations) are negligible since these processes are intrinsically non-classical \cite{rein}.  Furthermore, despite the slow velocities involved $(v_\perp \ll 1)$, quantum-mechanical spreading of the wavepacket representing the active electron can be neglected on the time scale during which charge transfer takes place.\\
To explore the approximate validity of some of these assumptions, quantum calculations have been performed, mostly for singly charged ions, where the reduction to an effective one-electron description is justified and where the density of states to be subtended in the calculations is limited. For simple metals, e.g. the collision of protons with an aluminum surface, perturbed atomic states of hydrogen can be calculated very accurately \cite{deutscher97,nord} using the potentials derived from ground-state density functional theory. Both the position of the resonance and its width, i.e. the lifetime and transition rate can be extracted from, for example, complex scaling \cite{nord} or stabilization calculations \cite{deutscher97}.\\
Despite the strong distortion of the atomic ground state, the bound state portion of the resonant wavefunction remains well-localized around the ion core  and traceable down to very small distances from the surface. Resonance positions remain well defined in the region where over-barrier transitions become possible (Fig. 6a). For large distances, the width of the resonance $\Gamma$ (Fig. 6b)
\begin{figure}[t]
\begin{center}
\epsfxsize=8cm   %width of figure - will enlarge/reduce the figures
\epsfbox{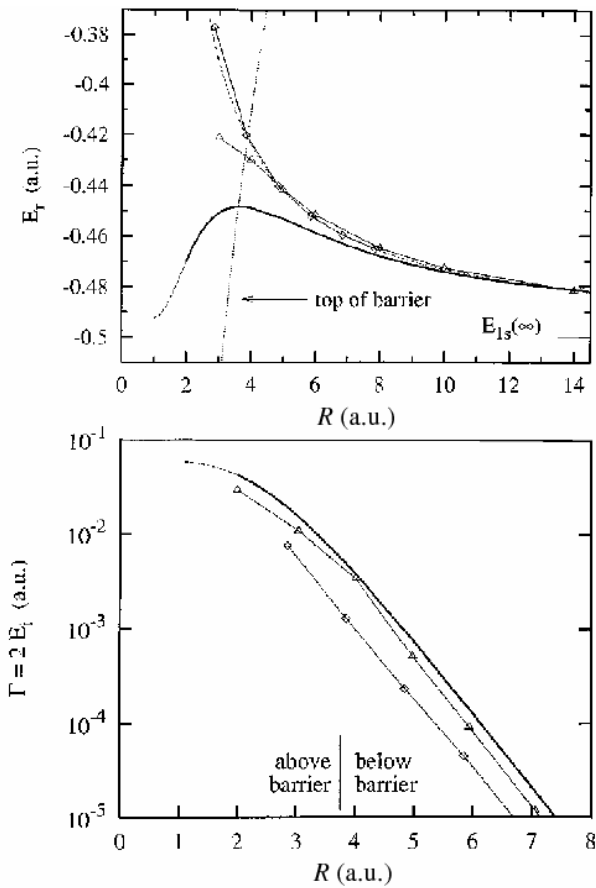}
\end{center}
\caption{
\label{fig:6a}
a) Position of the H(1s) resonance $[E_r=$Re$(E)]$ near an A1 as function of $R: -, \Delta$ and $\Diamond$ different complex-rotation calculations (ref. 25 and refs. therein), image shift formula $E_{1s} (\infty)$ + [1/4$(R-z_0)].$ Also shown is the energy of the top of the barrier as a function of $R$ (...).
%\label{fig:6b}
b) As in a) but for width of the $H(1s)$ resonance $[\Gamma = 2 $ Im$(E)]$ as a function of $R$.}
\end{figure}
decays approximately exponentially as expected for ``thick-barrier'' tunneling. At a critical distance $R_c \approx 3.6$ a.u., the resonance crosses the top of the barrier separating the atomic well from the jellium. As the 1s state becomes a broad ``over-barrier'' resonance, the width begins to saturate. This observation of well-defined, yet broadened over-barrier resonances supports the picture of over-barrier capture into broadened projectile states invoked in the COB model. For insulator surfaces, e.g. LiF as a prototype ionic alkali-halide surface, calculations of atomic energy levels in an environment of localized charges and low symmetry requires methods of quantum chemistry for large molecules at the level of self-consistent field \cite{bor,garcia} and beyond \cite{wirtz}. As contributions due to electron-electron correlations play an important role for the proper ordering of the covalent relative to the ionic levels, relevant for charge transfer, sophisticated methods such as the multi-reference configuration interaction (MR-CI) approach \cite{wirtz} are required. The position and width of avoided crossings (or more generally, conical intersections) near the LiF surface drastically change, when correlation effects are included. The relevant levels undergoing conical intersections have, so far, only determined for protons on LiF \cite{wirtz} and could provide a semiquantitative explanation for the ionization potential dependence of sputtering \cite{hayd}. For highly charged ions, a high density of accessible quasi-covalent levels with a multitude of conical intersections makes such calculations a formidable task. In fact, the very same reasons favor the applicability of the COB model. However, direct verification of the underpinning of the COB is still missing.\\
Solution of the time-dependent Schr\"odinger equations has, so far, only be attempted for one-electron problems, either employing coupled-channel methods \cite{deutscher98,borisov01} or wavefunction propagation or a grid  \cite{borisov00,chakra}. The latter was primarily used for the simulation of $H^-$ detachment, which can be reduced to an effective one-electron problem. The electronic ground state of the surface, e.g. of LiF, provides an adequate representation of the channel potential. Quantum calculations for this problems have yielded valuable insights into the role of the Madelung potential for the detachment near LiF \cite{borisov00} and the influence of the projected bandgap of $Cu$ \cite{chakra}. Extensions to HCI have not yet been attempted.\\ 
A potentially promising avenue for treating the time-dependent multi-electron dynamics is time-dependent density functional theory (TDDFT). As a matter of principle, time-dependent density functional theory \cite{runge} provides a highly efficient method to solve the time-dependent quantum many-body problem. It yields directly the time-dependent one-particle density $n(\vec{r},t)$ of the many-body system.\\
Within TDDFT, the time-dependent density is represented through the time-dependent Kohn-Sham spin-orbitals $\Phi_{\sigma,j} (\vec{r},t)$ as
\begin{equation}
\label{eq:10}
n(\vec{r},t)=\sum_{\sigma=\uparrow \downarrow} n_\sigma (\vec{r},t)=
\sum_{\sigma=\uparrow,\downarrow} \sum^{N_\sigma}_{j=1}
|\Phi_{\sigma,j}(\vec{r},t)|^2 \, ,
\end{equation}
where $N_\sigma$ denotes the number of electrons of spin $\sigma$. The one-particle spin-orbitals $\Phi_{\sigma,j} (\vec{r},t)$ evolve according to the time-dependent Kohn-Sham equation governed by the one-particle Kohn-Sham Hamiltonian
\begin{eqnarray}
\label{eq:11}
H^{KS}_\sigma [n_\uparrow, n_\downarrow] &=&-\frac{1}{2} \vec{\nabla}^2 + V_{ext}(\vec{r})+V(\vec{r},t)\\\nonumber
&& +V_H[n](\vec{r},t)+V_{xc}[n_\uparrow, n_\downarrow](x,t) \, ,
\end{eqnarray}
which includes the external one-particle potential, the Hartree potential and the exchange-correlation potential. The initial states $|\Phi_{\sigma,j}(t \rightarrow -\infty) \rangle = |\Phi_{\sigma,j}\rangle$ are the occupied Kohn-Sham orbitals of stationary ground state density functional theory (DFT).\\
In practice, however, applications are hampered by the lack of knowledge about the exact exchange-correlation potential $V_{xc}[n]$. Its form is a priori not known and a convergent algorithm for its calculation to increasingly higher degree of accuracy has not yet been proposed.
Moreover, the number of occupied orbitals needed to adequately represent the density of states of a surface of a simple metal is large $(N_\sigma \gtrsim 10^3)$, thus represent a major numerical challenge. This is most likely the reason why only few applications of TDDFT to ion-surface scattering \cite{shap,lem05} have been reported to date. A recent example for the time-dependent density fluctuation induced by a triply charged ion $(Q=3)$ in front of a jellium surface is shown in Fig.\ \ref{fig:7}.
\begin{figure}[t]
\begin{center}
\epsfxsize=8cm   %width of figure - will enlarge/reduce the figures
\epsfbox{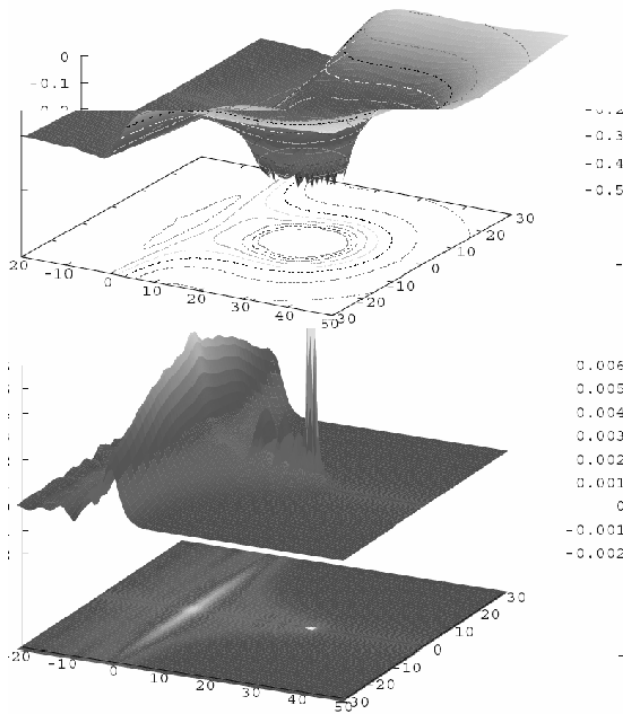}
\end{center}
\caption{
\label{fig:7}
Potential energy surface (top picture) and induced TDDFT density change (bottom picture) for a triply charged ion approaching a jellium surface with $r_s = 3$ and $W = 0.33$ a.u. for HCI-surface distance of $R=25 \, a.u.$.}
\end{figure}
At the surface the polarization charge density giving rise to the image potential can be observed. Simultaneously, the onset of charge transfer to the projectile becomes visible. At this large distance, $R=25$, capture still proceeds by tunneling which is reflected in the low electron density, $N_e=\int_V n d^3 r \ll 1$ when integrating over a volume enclosing the projectile. Only near $R_c$ for over barrier processes reaches $N_e \approx 1$.\\
A major conceptually difficulty is that, even if $n(\vec{r},t)$ would be exactly known, a read-out functional to extract occupation numbers of excited projectile and target states is still missing. Only very recently, some progress in the construction of a functional that allow to extract the S-matrix from the density, $S[n]$, has been made \cite{rohr}.

\subsection{Hybrid Classical-Quantum Simulations: Classical Transport Theory}
\begin{figure}[t]
\begin{center}
\epsfxsize=8cm   %width of figure - will enlarge/reduce the figures
\epsfbox{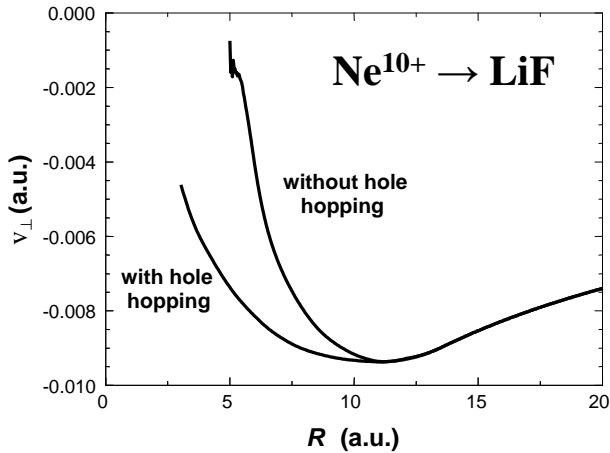}
\end{center}
\caption{
\label{fig:8}
Evolution of the average vertical velocity $v_z$ of a Ne$^{10+}$ ion with an initial energy $E_{kin}=1$ a.u. starting at a distance of 20 a.u. as a function of the distance from the surface. Solid line: with hole hopping; dashed line: hole hopping switched off.}
\end{figure}

Going beyond simple one-electron (or mean field) descriptions requires novel concepts. We have recently introduced a classical transport theory (CTT) which is based on a multi-particle Liouville master equation \cite{wirtz03}. It invokes four major ingredients: a) the explicit treatment of multi-electron processes by following the time evolution of the joint phase space density $\rho \left(\hbar, \vec{R},\{ P^{(P)} \}, \{P^{(T)}\}\right) $ that depends on population strings of $N$-electron states in the projectile $\{P^{(P)}\}$ and  $\{P^{(T)}\}$ target,
b) the usage of transition rates in the relaxation (or transport) kernel, that are derived from quantum calculations (mostly, first-order perturbation theory) wherever available, e.g. two-center Auger capture and deexcitation rates \cite{burg03}, and 
c) the embedding of these processes within the framework of a classical phase space transport simulation for the ion. The equation of motion of $\rho$ is of the form of a Liouville master equation
\begin{equation}
\label{eq:12}
\left( \frac{\partial}{\partial t}+\dot{\vec{R}} \vec{\nabla}_R-\frac{1}{M} \frac{\partial V_p}{\partial R} \; \vec{\nabla}_{\dot{\vec{R}}} \right)
\rho = \mathcal{R} \rho \, ,
\end{equation}
where the ``relaxation'' (collision) operator includes single and double particle-hole (de) excitation processes which represent resonant capture, resonant loss, hole hopping, ionization by promotion through the continuum, Auger capture, Auger deexcitation and autoionization which depend on both the local position of the ion, $\vec{R}$, and the population strings $\{p^{(p)} \}$ and $\{p^{(T)}\}$. A detailed discussion of the rates entering Eq. (\ref{eq:12}) is given in Ref. \cite{wirtz03,burg03}.
The effective projectile potential $V_p$ that governs the ionic motion will depend, in general, on the strings as well, i.e. $V_P\left(\vec{R}, \{P^{(P)}\}, \{P^{(P)}\} \right)$. 
Direct integration of the Liouville master equation (Eq. (\ref{eq:12})) appears to be extremely difficult in view of the large number of degrees of freedom involved. Here, the fourth ingredient,
d) solution by test particle discretization and a Monte Carlo sampling for ensembles of stochastic realizations of trajectories comes into play. We follow a large number of ionic trajectories with identical initial conditions for the phase space variables ($R,v_p)$ along an event - by - event sequence of stochastic electronic processes whose probability laws are governed by the rates of the underlying Liouville master equation. The probability for any process with transition rate $\Gamma^\alpha$ within a time interval $\Delta t$ to occur is determined by 
\begin{equation}
\label{eq:13}
W^\alpha(\Delta t)=1 -\textnormal{exp}(-\Delta t \Gamma^\alpha) \, .
\end{equation}
In order to decide which electronic transition (if any) takes place during the time period $\Delta t$, we use the rejection method for each of the distributions. At the same time the coordinate and velocity of the HCI are propagated in time according to a Langevin equation of motion.\\
The resulting population strings $\{P_n^{(P)}(t) \}_\mu$ and $\{P^{(T)} (t) \}_\mu$ for a single stochastic trajectory $\mu$ are discontinuous functions of time. After sampling a large number of trajectories, one obtains smooth ensemble averages representing solutions of Eq. (\ref{eq:12}). Earlier and less complete simulations of the neutralization scenario were given in Refs. \cite{win,burg95}. Variants of the present approach have been previously employed for energetic electron transport through solids \cite{burg90}, ion transport through nanocapillaries to be discussed below, and very recently for the interaction of strong laser fields with large clusters \cite{deiss}.\\ 
For illustration, we present an application of Eq. (\ref{eq:12}) to the interaction of Ne$^{10+}$ with a LiF surface in vertical incidence. We focus on the existence of the ``trampoline effect''. A microscopic trampoline effect was proposed by Briand et al. \cite{briand96} for insulators. As the HCI approaches the surface, the formation of a hollow atom is accompanied by the microscopic charge up of the surface in the vicinity of the impact region. As surface charges (i.e.holes) feature only a slow mobility, the ion may be repelled by the charge patch without actually touching down. The simulation of the average ion velocity (Fig.\ \ref{fig:8})
requires a simultaneous simulation of the random walk of the electronic holes on the $F^-$ sublattice which represent in this case the target strings $\{P^{(T)}\}$. The average velocity $v_\perp$ of the projectile remains always negative, meaning movement towards the surface. At distances larger than the critical distance for first electron capture, the projectile is accelerated by the self-image interaction. As electron capture begins to contribute, the acceleration still continues but is reduced because the charge state of the projectile and its image has decreased and because the repulsion due to holes generated by capture increases. At around 11 a.u., the hole repulsion starts to dominate over the image acceleration and the projectile slows down. The repulsive force can offset the image acceleration. However, it is, on the average, not strong enough to lead to a complete stop and to a reversal of the projectile {\it above} the surface. Only 2~\% of all trajectories are reflected at distances larger than 3 a.u. from the topmost layer and no turning point was observed at a distance larger than 3.5 a.u. Such small distances of closest approach correspond already to the fringes of the binary collision regime and imply an (almost) complete neutralization of the highly charged ion.\\
While we conclude that for a Ne$^{10+}$ vertically incident on an LiF surface, the trampoline effect, i.e., the above surface reflection leaving the ion in a multiply charged state, is absent, its occurrence for very high charge states where repulsion by slow holes should play a more prominent role remains an open question to be explored in the near future \cite{aumayr}.\\
\section{Ion Transport Through Nanocapillaries}
Nanocapillaries play currently a very prominent role in HCI-surface interaction well beyond the original goal to study the early stages of the hollow atom formation. The focus has shifted towards transport through capillaries, particularly of ions in their original charge states. For metallic capillaries trajectories of type 1 (see Fig.~\ref{fig:3}) transport promises new information on the stopping power (or friction force) at unprecedentedly large distances from the surface \cite{tokesi04}. For insulating capillaries a recently discovered ion guiding effect \cite{stolt} due to self-organized charge up suggests the opportunity to build ion-optical devices for forming and guiding nano-sized beams.\\
\subsection{Energy Loss in Metallic Nanocapillaries}
\begin{figure}[t]
\begin{center}
\epsfxsize=8cm   %width of figure - will enlarge/reduce the figures
\epsfbox{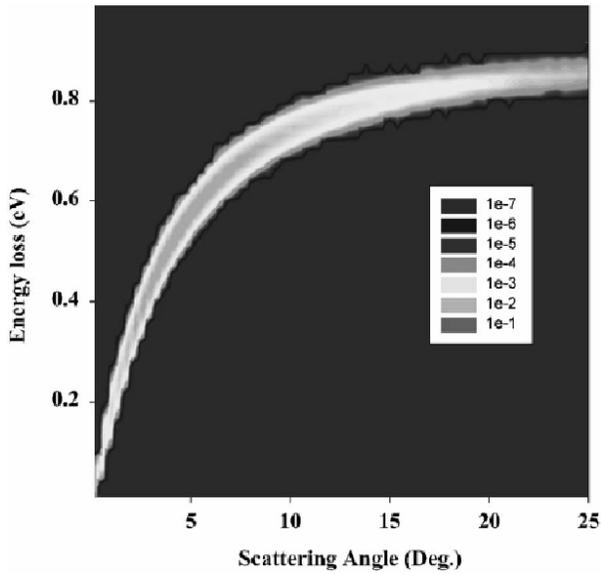}
\end{center}
\caption{
\label{fig:9}
2D correlation pattern between the energy loss and the scattering angle of Kr$^{30+}$ ions passing through an Ni nanocapillary at 2.5 eV/amu energy. The distance-dependent stopping power is calculated by the SRM (see text).}
\end{figure}

Calculation of the friction force for ions propagating parallel to the surface at large distances has remained a puzzle. Linear response (LR) theory for an electron gas within the framework of TDDFT yields an $R_z^{-4}$ distance dependence from the surface caused by particle-hole excitations \cite{liebsch}. The apparently much simpler approach of the socalled specular reflection model pioneered by Ritchie et al. \cite{ritchie} predicts a much stronger friction force $S$ decaying as $R_z^{-3}$. This difference is of crucial importance for transport of highly charged ions through nanocapillaries because of the high charge state $Q \gg 1, S \propto Q^2$ and the long interaction time during the propagation over mesoscopic distances $( \approx 1 - 10 \, \mu m)$. This puzzle was recently solved \cite{tokesi04} by noting that TDDFT-LR lacks the contribution of plasmon excitation of a jellium at large distances. In distant collisions, the long-wavelength or optical limit $(Q \rightarrow 0)$ is probed. Because of the lack of electron-phonon coupling in a jellium, the width $\gamma$ of the plasmon peak vanishes as $Q \rightarrow 0$. Consequently, plasmons can neither decay nor be excited. By correcting for the finite width at $\gamma (Q \rightarrow 0)$ within a modified TDDFT calculation, the proper large distance behavior could be restored. Fig.\ \ref{fig:9}
displays the correlated energy loss scattering angle $(\Delta E \theta)$ distribution of slow $Kr^{30+}$ ions penetrating a metallic Ni nanocapillary. The predicted energy loss is found to be sufficiently large as to be accessible by future experiments.
\subsection{Guiding through Insulating Capillaries} 
Capillaries through insulating foils (PET or ``Mylar'', \cite{stolt} and SiO$_2$ \cite{vikor}) have been studied in several laboratories \cite{kan,aumayr05}. Unexpectedly, considerable transmission probabilities for projectiles in their initial charge state were measured for incidence angles as large as $\approx 20^o$. Apparently, ions are guided along the capillary axis with a spread (FWHM) of $\Delta \theta_{out}$ of several degrees for mylar \cite{stolt} but close to geometric opening $\theta_0$ for SiO$_2$ \cite{vikor}. Keeping the initial charge state, contrary to the expected neutralization upon approach of the internal capillary surface, suggests that the ions bounce off the walls at distances larger than the critical distance $R_c \approx \sqrt{2Q/W}$ (Eq. (7)). We refer to this effect as a mesoscopic trampoline. Key to this process is the charging up of the internal insulator walls due to preceding ion impacts. Ion guiding through the capillary ensues as soon as a dynamical equilibrium of self-organized charge up by the ion beam, charge relaxation, and reflection is established.\\
A theoretical description and simulation of this process poses a considerable challenge in view of the widely disparate time scales simultaneously present in this problem:\\
The microscopic charge-up and hole transport due to the impact of individual ion impact takes place on a time scale of sub-$fs$ to $fs$ with a typical hole hopping time $\tau_h < 10^{-15}s$. The transmission time $\tau_t$ of a projectile ion through the capillary for typical ion energies of $\approx$ 200 eV/u is of the order of $\tau_t \approx 10^{-10} s$. Typical average time intervals $\overline{\Delta t}$ between two subsequent transmission (or impact) events in the same capillary are, for present experimental current densities of $nA/mm^2$ of the order of $\overline{\Delta t}\approx$ 0.1 s, and finally, characteristic (bulk) discharge times $\tau_b$ for these highly insulating materials, can be estimated from conductivity data to typically exceed $\tau_b \gtrsim 10^3$ s and can even reach days.\\
This multi-scale problem spans a remarkable 18 orders of magnitude. A fully microscopic ab initio simulation covering all relevant scales is undoubtedly out of reach. The method of choice is therefore a simulation based on the classical transport theory discussed above, modified such that the discharge characteristics deduced from data for macroscopic material properties of the nanocapillary material can be incorporated. Specifically, the bulk discharge time $\tau_b$ or bulk diffusion constant $D_b$ as well surface charge diffusion constant $D_s$ will be estimated from surface and bulk conductivity data for mylar \cite{ref:57}.\\
\begin{figure}[t]
\begin{center}
\epsfxsize=8cm   %width of figure - will enlarge/reduce the figures
\epsfbox{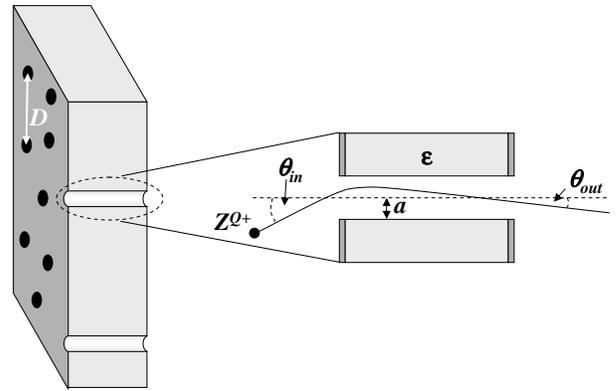}
\end{center}
\caption{
\label{fig:10}
Illustration of transmission through insulating nanocapillaries, schematically. Array of nanocapillaries oriented along the surface normal, inset: close-up of an individual capillary. An insulating  foil (PET) with dielectric constant $\varepsilon$ is covered on both sides with gold layers (dark shaded) preventing charge up of the target during experiment. Capillaries with radius $a$ = 50 nm and $L$ = 10$ \mu$m are typically $D$ = 500 nm apart. Projectiles enter and exit the capillary under angles $\theta_{in}$ and $\theta_{out}$ with respect to the capillary axis respectively. The capillary axis is either normal to the surface or Gaussian distributed with $\Delta \theta_\alpha \lesssim 2^o$ (FWHM).}
\end{figure}
The present approach represents a mean-field classical transport theory \cite{schiessl} based on a microscopic classical-trajectory Monte Carlo (CTMC) simulation for the ion transported, self-consistently coupled to the charge-up of and charge diffusion near the internal capillary walls. Initially, each ion impact at the surface deposits Q charges. The charged-up micro-patch will undergo surface diffusion with diffusion constant $D_s$ as well diffusion into the bulk with diffusion constant $D_b$. Bulk diffusion is extremely slow for highly insulating materials while the surface diffusion towards the grounded metallic layers (Fig.\ \ref{fig:10}) will be a factor $\cong $ 100 faster, thus governing the overall discharge process. Self-organized guiding sets in when a dynamical equilibrium between charge-up by a series of ion impacts at internal walls is established such that the electrostatic repulsion prevents further impacts and the ion is reflected at distances from the wall larger than the critical distance (Eq. \ref{eq:7}) from the surface. The wall forms then an effective mesoscopic ``trampoline'' for subsequent ions and guides the projectile towards the exit, as shown for a few sample trajectories in Fig.~\ref{fig:11}.
\begin{figure}[t]
\begin{center}
\epsfxsize=8cm   %width of figure - will enlarge/reduce the figures
\epsfbox{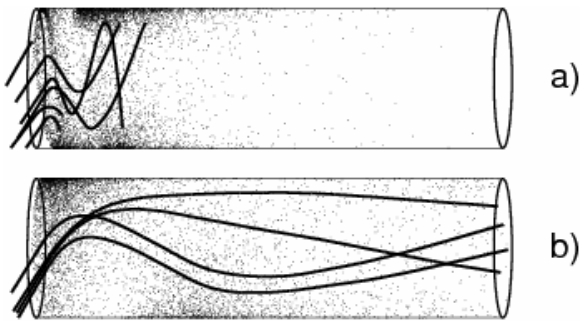}
\end{center}
\caption{
\label{fig:11}
Scatter plot of deposited charges in the interior of an individual capillary and resulting trajectories for $\theta_{in}=3^o$. a) zig-zag distribution leading to blocking for an unrealistic choice $(D_s = D_b)$; b) patch distribution leading to transmission for realistic values ($D_s=100 D_b)$.}
\end{figure}  
\\
A first quantitative comparison of the CTT simulation with experimental data can be made for the transmission probability as a function of the incident tilt angle $\theta_{in}$ relative to the capillary axis (Fig.~\ref{fig:12}).
Transmission occurs for $\theta_{in}$ well outside the geometric opening angle $\theta_0 \approx 0.5^0$ of the capillary. Theoretically predicted \cite{schiessl} efficiency in ion guiding agrees reasonably well with experimental data of Vikor et al. \cite{vikor}. Unlike the transmission function, the angular distribution of guided ions in the initial charge state is still not completely understood. The angular spread is, in part, determined by the spread in capillary axis distribution which is poorly known as well as by collective effects due to the charge-up of the ensemble of capillaries which can be viewed as a charged condenser \cite{schiessl}. The width of the angular distribution observed in some of the experiments exceeds the predicted width even if these additional sources of spreading are included. 
\begin{figure}[t]
\begin{center}
\epsfxsize=7cm   %width of figure - will enlarge/reduce the figures
\epsfbox{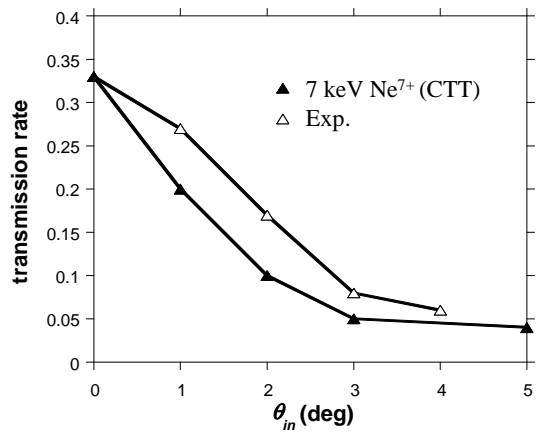}
\end{center}
\caption{
\label{fig:12}
Transmission function (transmission probability) as a function of angle of incidence $\theta_{in}$ relative to the mean capillary axis. Full symbols: present CTT, open symbols: experimental data: 7 keV Ne$^{7+}$ (Vikor et al. 58). Experimental transmission rates have been normalized to CTT results at $\theta_{in}=0^o$.}
\end{figure}
\section{Applications to Material Science}
The interaction of HCI with surfaces depends on the electronic structure and morphology of the surface. In turn, deposition of a large amount of potential energy onto an nanometer-sized impact zone leads to modifications of the surface. HCI surface collisions are therefore promising candidates for both localized surface diagnostics as well as surface modification and nanostructuring.
\subsection{Two-Dimensional Surface Magnetism}
\begin{figure}[t]
\begin{center}
\epsfxsize=7cm   %width of figure - will enlarge/reduce the figures
\epsfbox{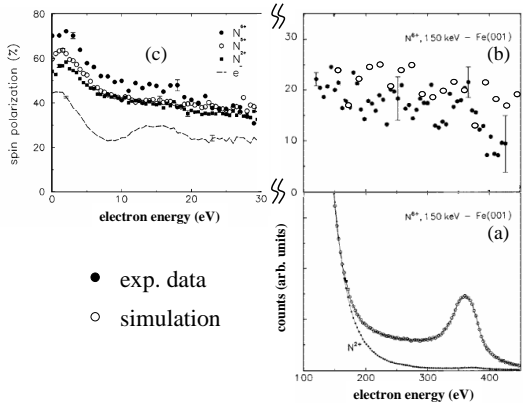}
\end{center}
\caption{
\label{fig:13}
Grazing incidence scattering $(\theta_{in}=1.5^o)$ of N$^{6+}$ at magnetized Fe(001) surface.
a) Exp. electron spectrum (Ref. 65) 
b) Exp. (65) and simulated (66) spin polarization of energetic electrons
c) Exp. spin polarization for different incident charge states at low energies.}
\end{figure}

(Anti) ferromagnetic ordering is a cooperative phenomenon and, thus, strongly dependent on the local environment to which atoms with a large magnetic moment are exposed to. One key parameter is the coordination number, the number of nearest neighbors. Surface and bulk atoms are subject to different environments, in particular, different coordination numbers. Magnetic ordering in the bulk (3D magnetism) may therefore be drastically different from surface (or 2D) magnetic ordering. Grazing incidence ion-surface scattering was very early recognized as an excellent tool to selectively probe the magnetic ordering of the topmost layer. Rau and collaborators \cite{rau} pioneered electron capture spectroscopy (ECS) which detects magnetization via spin polarization of captured electrons and the subsequent measurement of the polarization transferred to the nucleus via hyperfine interaction. Optical polarization measurements of excited states provide a complementary route \cite{zimny}.\\
Very recently, the investigation of 2D magnetism has been extended to multiply charged ions \cite{pfand,unipan}. Spin-polarized electron emission by $N^{6+}$ scattered at a Fe (001) surface has been used as a ``label'' for the electrons involved in the hollow atom formation. Electrons from the conduction band are, in general, spin polarized while core states giving rise to ``side feeding'' [9] show little to no polarization. Comparison of a recent COB simulation \cite{soll} with the polarization data \cite{pfand} shows reasonably good agreement for energetic electrons originating from KLL Auger transitions in $N^{6+}$ (Fig.~\ref{fig:13}).
The participating electrons originate primarily from the early stages of the hollow-atom formation and thus from the conduction band. A minor contribution comes from Auger capture into lower-lying states. The emitted electrons at lower energies have a multitude of sources, including slowed-down KLL electrons, lower-energy autoionization electrons, kinetic electron emission as well as secondary electron emission. While above E$\approx 50 $ eV transport and slowing down can account for the polarization observed, the remarkably high degree of polarization (up to $\approx 70 \%)$ at very low energies is, so far, unaccounted for and awaits a convincing explanation. A spin-filter effect \cite{kam} is a promising candidate but quantitative details remain to be worked out.
\subsection{Sputtering and Nanostructuring}
Conventional sputtering is driven by the kinetic energy deposited by, typically, singly charged ions (``kineting sputtering''). Increasing the energy deposition requires, thus, higher kinetic energies which, in turn, leads to deeper penetration and bulk modification. It was very early realized \cite{par} that slow HCI's promise an interesting alternative scenario for sputtering. The energy deposited is potential energy (``potential sputtering''). The penetration depth can be kept to a minimum of about one monolyer resulting in ``soft'' sputtering. The latter notion does not imply that the electronic or lattice structure to remain intact. Quite to the contrary, the first scenario suggested, the ``Coulomb explosion'' \cite{par} of insulator surfaces, suggested a violent distortion and restructuring of the surface, without, however, significant penetration of the projectile into the bulk. This original proposal for potential sputtering was plagued by several conceptual difficulties: Even in well-insulating materials, the mobility of holes, i.e. electronic charge carries, is too fast such that an extended region of high-density charge depletion cannot be maintained for a time interval of an electronic sputtering time $t_s \approx 500 fs$ necessary to convert potential energy into kinetic energy of target ions \cite{aumayr01}. If one neglects hole mobility, sputtering via Coulomb explosion would predominantly emit ionized target atoms with high kinetic energies $(\gtrsim 100 \, eV)$. Both of these predictions \cite{cheng} are at odds with several experiments. A variant to this scenario, Coulomb explosion of hydrogenated adsorbates has, however, been verified: the adsorption of hydrocarbons at surfaces results in the emission of protons which rapidly increases with the charge of the incident HCI as $\approx Q^4$. This has been explained \cite{burg96} in terms of $C-H$ bond breaking by the HCI and subsequent Coulomb explosion of the molecule. Here, the low mass of the proton and the orientation the hydrocarbon bond sticking out of the surface plays a crucial role in emitting protons sufficiently fast so that re-neutralization on the way out or reconstruction of a bond becomes unlikely.\\
\begin{figure}[t]
\begin{center}
\epsfxsize=8cm   %width of figure - will enlarge/reduce the figures
\epsfbox{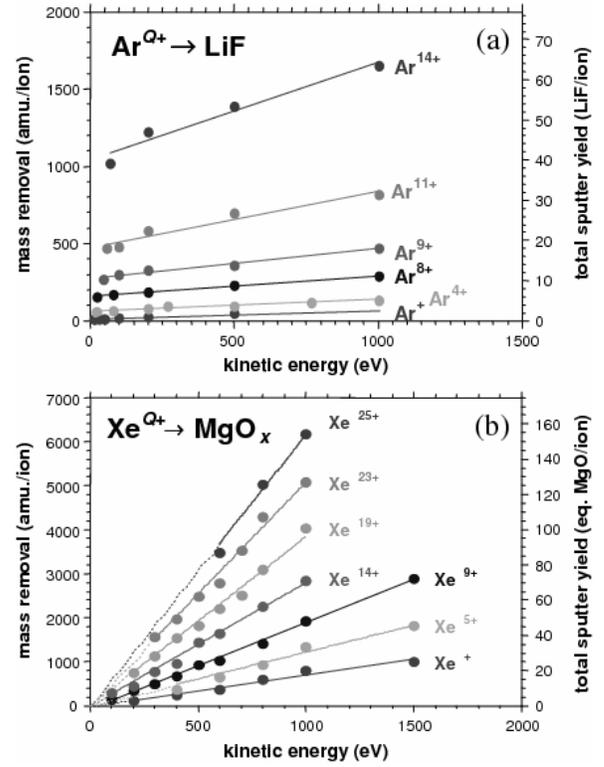}
\end{center}
\caption{
\label{fig:14}
Mass removal due to sputtering of (a) LiF and (b) of MgO$_x$ by highly charged ions as a function of ion impact energy. Left ordinate: in atomic mass units per incident projectile (as measured by the quartz crystal microbalance). Right ordinate: corresponding sputter yield (in molecules per incident ion). Solid lines for guidance only; dashed lines: extrapolation to zero kinetic energy (from refs. 12 and 71).}
\end{figure}
The simple Coulomb explosion picture has therefore be amended to yield a more complex scenario: central to a realistic sputtering mechanism is the trapping of the initially generated electronic defects, either by strong electron-phonon coupling to so-called ``self-trapped holes'' or ``self-trapped excitons'' or trapping by lattice defects generated by the incident projectile or already present. In either situation, the holes or, more generally, the electronic defects become transiently trapped so that potential energy can be converted within time $\tau_s$ into kinetic energy of target atoms. This conversion does not necessarily require ionic Coulomb interaction and thus allows for the emission of neutrals. A prominent example are color centers in alkali halides leading to copious emission of neutral Li and F atoms (Fig. 14a). Even at neglegible kinetic energy, a significant mass removal is achieved that is proportional to the potential energy, i.e. grows approximately with the charge $Q^2$. The primary production process of self-trapped holes and excitons has been understood, at least for singly charged ions, in terms of curve crossings (see Sect.~\ref{quantum} and refs. \cite{wirtz}). A more complex situation arises in materials where the electron-phonon coupling is weak, i.e. the trapping of primary electronic defects is insufficient to prevent rapid diffusion of holes. In these materials a different and more generally applicable mechanism has been identified, ``kinetically assisted potential sputtering (KAPS)'' \cite{hayd01}. The essence of this process, examples of which are shown in Fig. 14b, is the simultaneous presence of both kinetic energy and potential energy as prerequisite for large-scale ablation to take place. The efficiency of this process rapidly diminishes when either the kinetic energy or the charge state becomes small. This observation suggests that it is the kinetic energy of the incident projectile that causes defects (e.g. lattice dislocations) along its track which then serve as transient trapping centers for electronic defects caused by the accompanying potential energy. This interplay, which can be operative in a broad range of materials, makes potential energy induced sputtering a much more wide-spread phenomena.
%\clearpage
\subsection{Towards Nanostructuring}
\begin{figure}[t]
\begin{center}
\epsfxsize=8cm   %width of figure - will enlarge/reduce the figures
\epsfbox{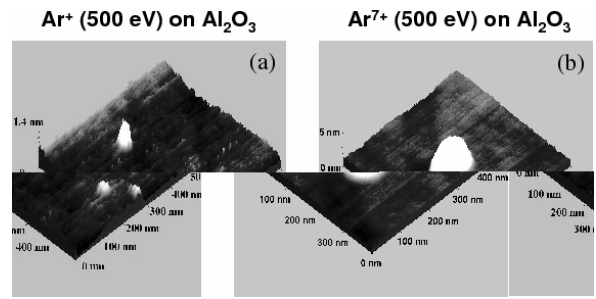}
\end{center}
\caption{
\label{fig:15} 
(a) UHV AFM contact mode image of sapphire (A1$_2$O$_3$, $c$-plane 0001) bombarded with 500 eV Ar$^+$ions. The defects are topographic features; all dimensions in nanometers, (b) as in (a) but bombarded with 500 eV Ar$^{7+}$ ions. Nanodefects induced by these ions with same kinetic but higher potential energy as compared to Ar$^+$, from ref. (74).}
\end{figure}

\begin{figure}[t]
\begin{center}
\epsfxsize=8cm   %width of figure - will enlarge/reduce the figures
\epsfbox{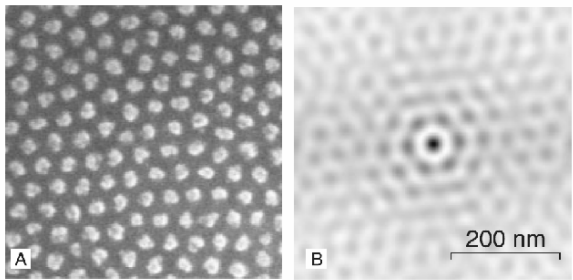}
\end{center}
\caption{
\label{fig:16}
SEM image of highly ordered cone-shaped dots on a (100) $GaSb$ surface formed by ion impact. (a) The extract of a SEM image and (b) the corresponding two-dimensional autocorrelation reveal the regularity and hexagonal ordering of the dots (from ref. 77).}
\end{figure}

The impact of a highly charged ion creates a strong dislocation in the surface (Fig.~\ref{fig:15})
which, contrary to naive expectation, is not necessarily a crater but can take on the shape of a ``blister''or ``hillock'' \cite{gebes,schneid}, as seen in AFM pictures. Clearly, the appearance on an AFM image may not accurately mirror the morphology of the defect as a clear-cut separation between microscopy and spectroscopy is difficult to achieve. In other words, an electronic defect may appear as a topographic defect \cite{megu}. Moreover, the size of the tip may be insufficient to distinguish the crater from its rim. Fig.~\ref{fig:15} also illustrates size dependence of the nanodefect on the charge. HCI are therefore excellent tools to inscribe nano-sized structures into the surface with the charge as control parameter of its size.\\
Future applications hinge on the availability of additional elements of control for nanostructuring. This is a largely unexplored area of research for potential sputtering. For kinetic sputtering, self-organization of defect structures have been analyzed \cite{brad} and, in part, experimentally verified \cite{bodek}. Bradley and Harper (BH) have introduced a phenomenological diffusion equation for the height variation $h(x,y)$ the $x-y$ plane, \cite{brad} 
\begin{equation}
\label{eq:14}
\frac{\partial}{\partial t} h(x,y,t)=-v_0 + \nu \nabla^2 h(x,y,t) - D \nabla^2  \left(\nabla^2 h(x,y,t) \right) \, ,
\end{equation}
where $v_0$ is the constant ablation rate, $\nu$ is the (negative) surface tension while $D$ is the diffusion rate for the surface curvature. Eq. (\ref{eq:14}) applies to normal incidence. Additional terms with odd powers of the derivate would appear for oblique incidence. Key to nanostructuring is the negative surface tension which describes, on a phenomenological level, the tendency towards spontaneous roughening of the surface under ion impact, i.e. randomly appearing depressions are deepened as they become more susceptible to further ablations than nearby hills. The microscopic justification for negative surface tension relies on the spike-like energy deposition pattern for kinetic sputtering. The balance of curvature diffusion which tends to smooth the surface and kinetic roughening results in self-organized nanostructuring. Ordered wavelike patterns have been predicted for oblique incidence. For normal incidence, a near-ordered hexagonal spatial correlation pattern in $h(x,y)$ has been recently observed \cite{bodek}, clearly pointing towards self-organized ordered nanostructuring (Fig.\ \ref{fig:16}).
However, such a pattern formation has not been theoretically accounted for in detail. Moreover, translating these observations to potential sputtering is not straight-forward. Its primary energy deposition is expected to be markedly different from spikes along tracks. The challenge for the future is thus to develop a theory in analogy to the BH approach Eq. (\ref{eq:14}) for potential sputtering, in particular to explore the existence (or absence) of the negative surface tension and to improve the BH equation such as to account for ``crystallization'' of defects in ordered hexagonal patterns.\\
 
Work supported by the Austrian Fonds zur F\"orderung der wissenschaftlichen Forschung under proj.\ nos.\ FWF-SFB016 ``ADLIS'' and P17449-N02, by the EU under contract No. HPRI-CT-2001-50036, by the Hungarian Scientific Research Found: OTKA Nos. T038016, T046454, the grant ``Bolya'' from the Hungarian Academy of Sciences, and the TeT Grant No. A-15/04. 

\end{document}